\documentclass[a4paper]{article}
\usepackage{INTERSPEECH2021}
\usepackage{subcaption}
\usepackage{makecell}
\usepackage{fixltx2e}
\usepackage[skip=3pt]{caption} 
\usepackage{multirow}
\usepackage{float}
\usepackage{subcaption}
\usepackage{tipa}
\usepackage{cite}
\usepackage{amsmath,graphicx}
\usepackage[table,xcdraw]{xcolor}
\usepackage{bm}

\title{Automatic Detection of Speech Sound Disorder in Child Speech Using Posterior-based Speaker Representations}

\name{Si-Ioi Ng$^1$, Cymie Wing-Yee Ng$^2$, Jiarui Wang$^1$, Tan Lee$^1$}
\address{
  $^1$Department of Electronic Engineering, The Chinese University of Hong Kong \\
  $^2$Department of Chinese and Bilingual Studies, The Hong Kong Polytechnic University
}

\email{siioing@link.cuhk.edu.hk, tanlee@ee.cuhk.edu.hk}

\begin{document}

\maketitle

\begin{abstract}


This paper presents a macroscopic approach to automatic detection of speech sound disorder (SSD) in child speech. Typically, SSD is manifested by persistent articulation and phonological errors on specific phonemes in the language. The disorder can be detected by focally analyzing the phonemes or the words elicited by the child subject. In the present study, instead of attempting to detect individual phone- and word-level errors, we propose to extract a subject-level representation from a long utterance that is constructed by concatenating multiple test words. The speaker verification approach, and posterior features generated by deep neural network models, are applied to derive various types of holistic representations. A linear classifier is trained to differentiate disordered speech in normal one. On the task of detecting SSD in Cantonese-speaking children, experimental results show that the proposed approach achieves improved detection performance over previous method that requires fusing phone-level detection results.
Using articulatory posterior features to derive i-vectors from multiple-word utterances achieves an unweighted average recall of 78.2\% and a macro F1 score of 78.0\%.

\end{abstract}
\noindent\textbf{Index Terms}: child speech, speech sound disorder, speaker representation, articulatory feature, speech attributes.

\section{Introduction}\label{intro} 

Speech sound disorder (SSD) is a common type of communication disorders affecting 2.5\% - 13.8\% of children aged below 8 \cite{kim2017prevalence,eadie2015speech,karbasi2011prevalence, wren2016prevalence}.
When growing up, children are expected to master the language's speech sounds in stages and be able to self-correct the pronunciation mistakes. 
A significant portion of children may encounter persistent difficulties in correctly producing certain speech sounds after the expected stage of acquisition. 
The symptom is one diagnostic attribute of SSD in clinical speech therapy. Untreated children with SSD are prone to unsatisfactory social and educational outcomes \cite{daniel2017children, hitchcock2015social}. Early diagnosis and intervention are necessary for effective treatment and rehabilitation. At present, 
clinical diagnosis of SSD is carried out by qualified speech and language pathologists (SLP). The evaluation of the speech sound inventories of children reveals the presence of SSD.

To timely identify children with SSD and refer to SLP for intervention, automated detection of speech disorder is considered a highly desirable approach for assessment and/or screening a large population of children.
Detection of SSD is commonly formulated as a task of distinguishing disordered speech sounds from typical ones at phoneme-level, based on acoustic speech signals.
Siamese neural networks were adopted to contrast hypothetically disordered consonant segments with typical ones \cite{wang19n_interspeech, ng20b_interspeech}. 
Posterior features were derived from automatic speech recognition systems to facilitate mispronunciation detection in disordered child speech \cite{hair2021assessing, shahin2019anomaly}.
Considering that consonant error could alter the acoustical characteristics of its neighbouring vowel, 
our recent work proposed to detect consonant errors from consonant-vowel speech segments \cite{ng21_interspeech}. 
The phoneme-level error detection requires automatic time alignment of the target phoneme segments. 
Time alignment could be inaccurate for disordered speech recorded in naturalistic communication scenarios. 
Inaccurate segments would affect the efficacy of model training and the detection performance.

Taking a different perspective, detection of SSD can also be formulated as a problem of measuring the overall goodness of the child's speech production, without going into specific parts of a long utterance.
The reliability concern of segment boundaries is thus bypassed.
Paralinguistic features defined in the extended Geneva Minimalistic Acoustic Parameter Set (eGeMAPS) represent the overall acoustic characteristics of a speech utterance using statistical functionals \cite{eyben2015geneva}. 
They were successfully applied to detect atypical speech, including disordered speech with phonological and articulation problems \cite{shahin2019automatic}.
The approach of speaker verification (SV) was also investigated in classifying disordered speech  \cite{laaridh17_interspeech, moro2020using, laaridh18b_interspeech, quintas20_interspeech}. 
In \cite{kothalkar2018automatic, kothalkar2018fusing}, SSD subjects were identified by a text-dependent SV system trained on single-word utterances.
The word-level detection results were fused to give a subject-level score as an overall assessment of word articulation. 

Spectral features, e.g. Mel-frequency cepstrum coefficients (MFCCs), have been routinely used to extract speaker representations. Posterior features derived from deep neural network (DNN) classifiers are closely related to the linguistic properties of speech \cite{behravan2015vector, d2014extended}. 
They particularly characterize the articulation aspect of speech sounds. Compared to conventional spectral features, posteriors are believed to convey more pertinent information that can help detect atypical speech sounds \cite{wang19n_interspeech, shahin2019anomaly, hair2021assessing}. We expect that using speaker representations extracted from posterior features would effectively differntiate disordered children from typically developing (TD) ones.

In the present study, we investigate the possibility of transforming the word articulation test for SSD detection into a holistic assessment of multiple word utterances. Prescribed target words are grouped together, 
and the subject-level judgement is given based on the overall discrepancy between disordered speech and typical speech. The detection does not require precisely locating, analysing and classifying individual phones/words.
Towards capturing the coarse-grained speech characteristics of long utterances, speaker representations such as i-vectors and x-vectors have been investigated  \cite{dehak2010front,snyder17_interspeech}. 
Using word-level utterances ensures that every single error in the pronunciation can be covered, and speech duration would not be too short for feature extraction \cite{mclaren2010experiments}.
In view of the limited amount of child speech data, we need to determine the type of speaker representation that can perform well in the low-resource scenario.
We also investigate the use of knowledge-driven posteriors in the extraction of speaker representations. 
Grouping the phonemes based on speech attributes is adopted in the 
training of posterior extractors, as speech attributes were shown to provide diagnostic information about SSD symptoms \cite{wang19n_interspeech, shahin2019anomaly}. 
We will compare the proposed speaker representations for subject-level SSD detection with phone-level and word-level approaches.

\section{Child Speech Database}

This research is focused on SSD in Cantonese-speaking pre-school children. 
Cantonese is a major Chinese dialect widely spoken by millions of people in Hong Kong, Macau, Guangdong Provinces of Mainland China, and many overseas Chinese communities. It is a monosyllabic and tonal language. Each Chinese character is pronounced as a single syllable carrying a lexical tone. There are $19$ initial consonants, $11$ vowels, $10$ diphthongs, $6$ final consonants and $6$ lexical tones \cite{zee1999acoustical,bauer2011modern}.

\begin{table}[t]
\caption{Speech attributes of Cantonese.}
\label{attribute_map}
\centering
\resizebox{0.90\linewidth}{!}{%
\begin{tabular}{|c|c|c|}
\hline
\rowcolor[HTML]{EFEFEF} 
\textbf{Category}                     & \textbf{Feature} & \textbf{Phoneme}                                                       \\ \hline
                                      & Plosive            & p p\textsuperscript{h} t t\textsuperscript{h} k k\textsuperscript{h}  k\textsuperscript{w} k\textsuperscript{wh}        \\ \cline{2-3} 
                                      & Nasal              & \textipa{m n N}                                                \\ \cline{2-3} 
                                      & Affricate          &     \textipa{ts ts\textsuperscript{h}}                                                   \\ \cline{2-3}
                                      & Fricative          & \textipa{s f h}                                                           \\ \cline{2-3}
                                      & Glide              & \textipa{j w}                                                          \\ \cline{2-3} 
\multirow{-6}{*}{\textbf{Manner}}     & Liquid             & \textipa{l}                                                                  \\ \hline
                                      & Aspirated          & p\textsuperscript{h} t\textsuperscript{h} k\textsuperscript{h}  k\textsuperscript{wh} ts\textsuperscript{h}                      \\ \cline{2-3} 
   \multirow{-2}{*}{\textbf{Aspiration}}                               & Unaspirated        & \textipa{p t k k\textsuperscript{w} ts} 
  \\ \hline
                                      & Alveolar           & \textipa{t t\textsuperscript{h} ts} \textipa{ts\textsuperscript{h} s j}                         \\ \cline{2-3} 
                                      & Lateral            & \textipa{l}                                                                   \\ \cline{2-3} 
                                      & Labial             & \textipa{p p\textsuperscript{h} w m}                      \\ \cline{2-3} 
                                      & Velar              & \textipa{k k\textsuperscript{h} N}                                                 \\ \cline{2-3} 
                                      & Labio-Velar       & \textipa{k\textsuperscript{w} k\textsuperscript{wh}}                                                        \\ \cline{2-3} 
                                      & Labio-dental      & \textipa{f}                                                                 \\ \cline{2-3} 
\multirow{-7}{*}{\textbf{Place}}      & Vocal              & \textipa{h}                                                                  \\ \hline \hline
\textbf{Vowel/Semi-vowel} && \textipa{a: i: E: e} \textipa{œ: œ O: o u: y:} \textipa{5 I 8 U } \\ \hline
\end{tabular}%
}
\end{table}

Experiments on subject-level SSD detection are carried out with 
a large-scale child speech database named CUCHILD \cite{ng20_interspeech}. The database contains speech data from $1,986$ kindergarten children aged $3$ - $6$ in Hong Kong. All speakers use Cantonese as their first language.
The speech materials consist of $130$ Cantonese words of $1$ to $4$ syllables in length. These words cover all Cantonese consonants and vowels. 
Each subject in CUCHILD was formally assessed with the Hong Kong Cantonese Articulation Test (HKCAT) \cite{cheung2006hong}. About $230$ children in the database were found to have SSD. Most of these children committed speech sound errors more frequently in the initials consonants than vowels. 

We select speech data from $415$ subjects in CUCHILD, including $265$ typically developing (TD) children and $150$ disordered.
48\% of the TD and 61\% of the disordered children are aged 3-4. 
Speech sound errors made by the disordered speakers are carefully annotated by four student clinicians from speech therapy programmes. 
Given the heterogeneous acoustic environment of speech recording, for some subjects, a few spoken word items could not be located or were contaminated by background noise. 
$125\pm14$ word utterances are available for each speaker. For each disordered speakers, $45\pm25$ word utterances are annotated as having speech sound errors. 5-fold cross-validation experiments are carried out. 80\% of the TD and disordered speakers are used in model training, i.e. about $5.5$ hours of training data in each fold. Speech data from the training speakers does not appear in the test set.

\section{Proposed System and Feature Design}\label{proposed_features}
\subsection{System Architecture}
\begin{figure}[t!]
  \setlength\belowcaptionskip{-0.8\baselineskip}
  \centering
  \includegraphics[width=0.96\linewidth]{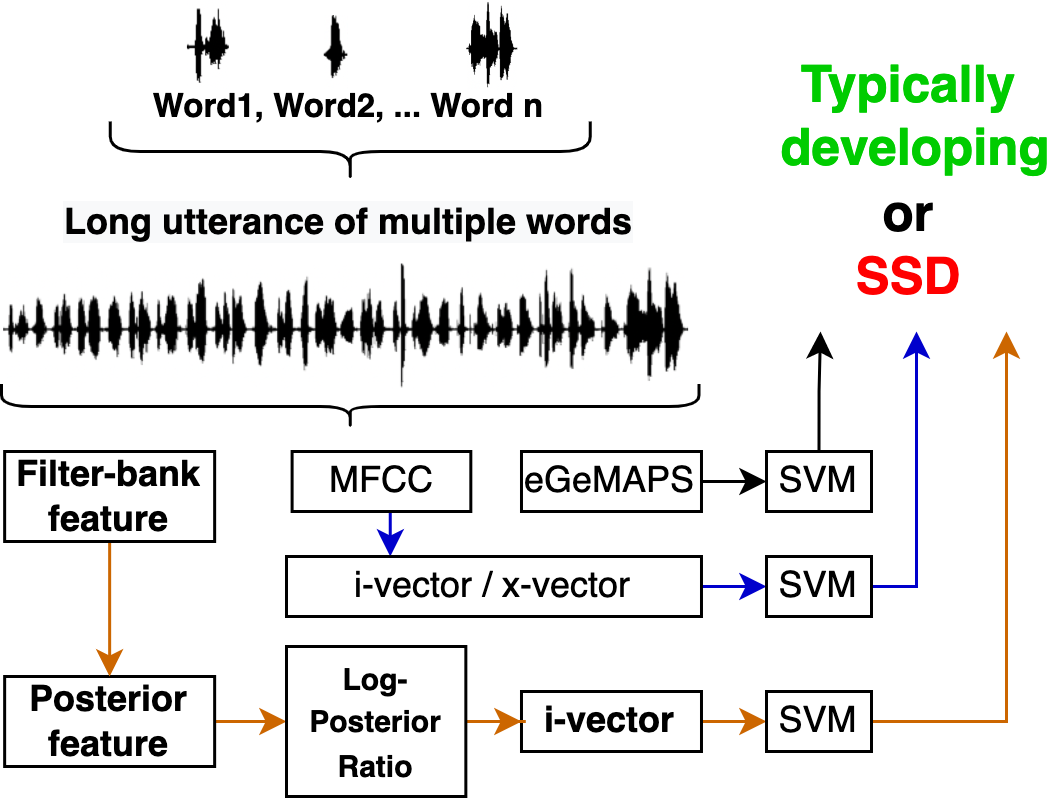}
  \caption{Subject-level SSD detection system.} 
  \label{fig:system_design}
\end{figure}

In the standardized assessment of SSD for young children, a set of designated test words are used for all subjects. The test words are selected purposely based on linguistic and clinical knowledge to cover the target speech sounds (phonemes) to be evaluated \cite{cheung2006hong, goldman2015}. 
Our proposed system for subject-level SSD detection is shown as in Figure \ref{fig:system_design}.
For each child subject, 
speech segments that contain the test words are concatenated to construct a long utterance of multiple words. Front-end feature processing involves the computation of filter-bank features, MFCCs or eGeMAPS. 
I-vector or x-vector is derived from the MFCCs as a speaker-level feature representation.
The i-vector can also be obtained from the posterior features generated by a DNN.
A support vector machine (SVM) is trained on the i-vector, x-vector or eGeMAPS to determine if the subject is TD or disordered.

\subsection{I-vector and x-vector}
I-vector is a fixed-dimensional representation of variable-length utterance from a speaker. Consider an utterance $i$. It is modeled by a super-vector, defined as 
$\bm{\mu}_{i} = \bm{m} + \bm{T}\bm{w}_{i}$, where  $\bm{m}$ is the mean vector obtained from Gaussian Mixture Model - Universal Background Model (GMM-UBM). The GMM-UBM is the speaker-independent component trained on all speakers.
$\bm{T}$ and $\bm{w_{i}}$ are the speaker-dependent components.
$\bm{T}$ is the total variability matrix and $\bm{w_{i}}$ is the latent factor, known as the i-vector.
Training of i-vector extractor is implemented by the expectation maximization (EM) algorithm. 

x-vector is extracted from a DNN model \cite{snyder17_interspeech}. A time-delay neural network (TDNN) architecture is adopted and trained with a cross-entropy objective function. Training of the TDNN requires splitting an utterance into segments. x-vectors of the segments are computed by aggregating the hidden representations of the TDNN via mean and variance pooling. They are averaged to derive the subject-level x-vectors.

I-vector and x-vector were applied to detecting speech-related diseases, e.g. dysarthria \cite{laaridh17_interspeech}, Parkinson's disease \cite{moro2020using}, and oral cancer \cite{laaridh18b_interspeech, quintas20_interspeech}. 
It was common to train the extractors using out-of-domain corpus, e.g. from healthy adult speakers.
In the present study, apart from using out-of-domain data, we attempt to use in-domain child speech data,  which incorporates disordered speech in training.


\subsection{Posterior feature for i-vector extraction}

Consonant substitution is one of the major types of speech sound errors in child speech. The errors are usually described as specific categories of substitution patterns. For example, the Cantonese initial consonant /\textipa{t}/ being substituted by  /\textipa{k}/ is known as the ``backing'' error.  /\textipa{ts\textsuperscript{h}}/ substituted by /\textipa{s}/ is described as a ``de-affrication'' error. 
These errors correspond to the change of speech attributes, i.e., the manner and place of articulation, and/or the aspiration status. In the example of /\textipa{t}/\textrightarrow/\textipa{k}/, the place of articulation shifts from bilabial to alveolar, while the example of /\textipa{ts\textsuperscript{h}}/\textrightarrow/\textipa{s}/ is due to change in the manner of articulation from affricate to fricative. 
The grouping of phonemes based on speech attributes is listed in Table \ref{attribute_map}. The Cantonese phonemes are represented in the international phonetic alphabet (IPA) symbols.

Posterior features derived from a DNN based speech attribute/phone classifier provide information related to the contrast and similarity between phonemes. 
The classifier adopts the architecture of bi-directional gated recurrent unit (Bi-GRU). The inputs to the Bi-GRUs are filter-bank features. 
Posteriors are extracted from the Bi-GRU output at each time frame using a fully connected layer. 
The posteriors are transformed into log-posterior ratio (LPR), which is defined as 
\begin{equation} \label{LPR}
\begin{split}
\text{LPR}_{i} = \log (\dfrac{p_{i}}{1-p_{i}}), 
\end{split}
\end{equation}
where $i=1,2,...,N$, with $N$ being the total number of phone/articulatory classes. $p_{i}$ represents the posterior probability of the $i$-th class \cite{diez2014projection}. Subsequently, the LPRs are used for the i-vector modeling.

\subsection{Paralinguistic features}
The eGeMAPS is a minimal set of paralinguistic parameters measured on speech signals \cite{eyben2015geneva}. A collection of low-level descriptors (LLDs) are computed to characterize the speech signal's frequency, energy, and spectral aspects at frame-level. The LLDs include acoustic measures such as pitch, formant frequency, MFCC, shimmer, jitter and spectral energy, etc. 
An 88-dimensional feature vector is derived by applying statistical functionals to the LLDs. Extraction of eGeMAPS features is implemented by the OpenSmile toolkit \cite{eyben2010opensmile}.

\section{Fusing Phone/Word-level Results}

Conventionally, SLPs observe the child's speech sound errors in specific parts of the test words. The results are recapitulated for diagnosis of the child. Our proposed approach based on speaker-level representation is compared with methods in which word-level or phone-level error detection is done first on individual test words and subsequently combined to obtain the subject-level decision \cite{ng21_interspeech, shahin2019automatic, kothalkar2018automatic}. 

Phone-level detection is implemented based on our previous work in  \cite{ng21_interspeech}. 
A DNN model is trained on the classification task to extract fixed-dimension embeddings of consonant-vowel segments.
For detection, each test embedding $\boldsymbol{x_{test}}$ is compared with the reference embeddings $\boldsymbol{x_{ref}}$ from TD speakers in a pairwise manner. 
Pairs of embeddings from TD speech are generated to train a logistic regression (LR) classifier based on the similarities given by $||\boldsymbol{x_{test}}-\boldsymbol{x_{ref}||_{1}}$.
The classification output tells whether the consonant-vowel segments in the test-reference pair are different. 
For each speaker, the accuracy score of detecting each initial consonant is stacked to derive a speaker-level representation that describes the child's performance on the articulation test.
A SVM is trained on the representations to classify the test subject as TD or disordered.

For word-level detection, i-vector, x-vector, or eGeMAPS can be extracted from single-word utterances. These representations are similar to the speaker-level representations introduced in Section \ref{proposed_features}, except that one representation would be obtained from each test word. Training labels of word-level representations are inherited from the speaker-level diagnostic labels, i.e. TD or SSD. An LR classifier is trained on the word-level representations. 
The majority of word-level classification results determines the subject-level result.

\section{Experimental Set-up}

\subsection{I-vector and x-vector extraction}

The i-vector and x-vector systems are implemented with the Kaldi toolkit \cite{povey2011kaldi}. 
Four i-vector systems are evaluated in our experiments. Training of the first system follows the approach in \cite{kothalkar2018automatic} using  MFCCs and single-word utterances.    
The other three systems are trained on long utterances of multiple words for extracting speaker-level representation. The input features are MFCCs, speech attribute LPR and phone LPR, respectively.
The i-vector systems use 256 mixture components, and the dimension of i-vectors is 100. 


Three different x-vector systems are trained and compared with the i-vector systems. 
The network architecture in \cite{snyder17_interspeech} is used with that the size of hidden layers reduced by half. 
The generated x-vector has a dimension of $256$. 
The first x-vector system is built using an out-of-domain corpus called CUSENT, which contains about $20$ hours of Cantonese speech  from $76$ adults speakers \cite{lee2002spoken}. The network is trained with speaker labels.
The other two x-vector systems are trained with child speech, using long utterances of multiple words, and single-word utterances, respectively.
When single-word utterances are used, discriminative training of x-vector extractors is carried out with the speaker labels. When utterances of multiple words are used, diagnostic labels (SSD or TD) are used to capture the difference between different TD and disordered speakers.

\subsection{Posterior and embedding extractors}

The phone-level or speech-attribute posterior extractors and the consonant-vowel embedding extractors all use the same neural network model with 3 layers of Bi-GRUs. The input acoustic features are 80-dimensional filter-bank features extracted every 0.01 second, and mean and variance normalised. Training speech data consist of consonant-vowels segments extracted obtained by forced alignment. 
The embedding and posterior extractors are trained as a softmax-based multi-class classification task with cross-entropy objective functions. 
The number of classification targets for phone and consonant-vowel units is $33$ and $173$, respectively.  
Multi-task training is carried out on 16 speech attributes for speech attribute posterior extractors.
The Adam optimizer is applied with a learning rate of $0.0001$ for the posterior extractors and $0.001$ for the embedding extractors \cite{KingmaB14}. The posterior extractors are built using the Phonet toolkit \cite{vasquez2019phonet} toolkit. The consonant-vowel embedding extractors are implemented by PyTorch \cite{paszke2019pytorch}.

\subsection{Back-end classifier for SSD detection}
SVM and LR classifiers for phone-, word- and subject-level detection are trained on speaker-level representations or paralinguistic features using sklearn \cite{pedregosa2011scikit}.
We assign positive label `1' to the representations of disordered speakers and negative label `0' to the TD ones.
The SVM uses a linear kernel. The regularization parameter of SVM and LR is $1.0$. 


\begin{table}[t!]
\centering
\caption{Classification performance of embedding and posterior extractors.}
\resizebox{0.90\linewidth}{!}{

\begin{tabular}{|c|c|c|} 
\hline
\textbf{Extractor}                & \textbf{Metric}        & \textbf{Performance}  \\ 
\hline
Consonant-vowel                   &       \multirow{2}{*}{Accuracy}                 & 88.0 (0.9)            \\ 
\cline{1-1}\cline{3-3}
Phone                             &                        & 70.8 (1.7)            \\ 
\hline
\multirow{2}{*}{Speech attribute} & Max. UAR (Vocal)       & 95.4 (1.0)            \\ 
\cline{2-3}
                                  & Min. UAR (Velar)       & 83.1 (0.6)            \\
\hline
\end{tabular}
}
\label{extractor_performance}
\end{table}

\begin{table}[t!]
\centering
\caption{Performance of subject-level SSD detection using  the proposed approach.}
\resizebox{\linewidth}{!}{
\begin{tabular}{|c|c|c|c|c|} 
\hline
\multirow{2}{*}{\textbf{Method}}                                                        & \multirow{2}{*}{\begin{tabular}[c]{@{}c@{}}\textbf{Dim.}\\\textbf{Reduction}\end{tabular}} & \textbf{\textbf{Classifier}} & \multirow{2}{*}{\textbf{UAR}}                  & \multirow{2}{*}{\textbf{Macro F1}}              \\ 
\cline{3-3}
                                                                                        &                                                                                            & Subject             &                                                &                                                 \\ 
\hline
eGeMAPs                                                                                 & \multirow{3}{*}{-}                                                                         & \multirow{3}{*}{SVM}         & 65.1 (5.3)                                     & 63.9 (5.1)                                      \\ 
\cline{1-1}\cline{4-5}
x-vector~                                                                               &                                                                                            &                              & 65.7 (6.2)                                     & 65.2 (6.1)                                      \\ 
\cline{1-1}\cline{4-5}
\begin{tabular}[c]{@{}c@{}}x-vector\textsubscript{(CUSENT)}\end{tabular}                             &                                                                                            &                              & 68.2 (5.9) & 67.1 (7.0)  \\ 
\hline\hline 
\multirow{2}{*}{\begin{tabular}[c]{@{}c@{}}i-vector\textsubscript{(CUSENT)}\end{tabular}}            & -                                                                                          & \multirow{8}{*}{SVM}         & 69.9 (3.6)                                     & 68.6 (4.0)                                      \\ 
\cline{2-2}\cline{4-5}
                                                                                        & LDA                                                                                        &                              & 72.0 (5.8)                                     & 71.3 (5.3)                                      \\ 
\cline{1-2}\cline{4-5}
\multirow{2}{*}{i-vector}                                                               & -                                                                                          &                              & 71.7 (6.3)                & 71.7 (6.3)                 \\ 
\cline{2-2}\cline{4-5}
                                                                                        & LDA                                                                                        &                              & 72.5 (7.3)                & 72.4 (7.7)                \\ 
\cline{1-2}\cline{4-5}
\multirow{2}{*}{\begin{tabular}[c]{@{}c@{}}i-vector\\~(\it{phone LPR})~\end{tabular}}            & -                                                                                          &                              & 74.5 (7.9)                                     & 73.8 (6.5)                                      \\ 
\cline{2-2}\cline{4-5}
                                                                                        & LDA                                                                                        &                              & 76.2 (6.5)                                     & 76.0 (6.6)                                      \\ 
\cline{1-2}\cline{4-5}
\multirow{2}{*}{\begin{tabular}[c]{@{}c@{}}i-vector \\(\it{speech attribute LPR})~\end{tabular}} & -                                                                                          &                              & 76.9 (5.3)                                     & 76.5 (5.6)                                      \\ 
\cline{2-2}\cline{4-5}
                                                                                        & LDA                                                                                        &                              & \textbf{78.2 (4.6)}                            & \textbf{78.0 (4.9)}                             \\
\hline
\end{tabular}
}
\label{subject_level}
\end{table}

\section{Results and Discussion}

The classification performance of the embedding and posterior extractors is evaluated on the speech data from TD speakers in each cross-validation fold. Results are reported in Table \ref{extractor_performance} in terms of classification accuracy and unweighted average recall (UAR). 
Overall, the extractors produce a fairly satisfactory performance in classifying the consonant-vowels, phones and speech attributes. The embeddings or posterior features extracted from
child speech utterances are deemed applicable and effective to subject-level SSD detection.

The 5-fold cross-validation results of subject-level SSD detection using the proposed approach are given in Table \ref{subject_level} in terms of UAR and the macro F1 score. 
With long utterances of multiple words, the i-vector approach surpasses x-vector and eGeMAPS. 
The i-vector approach remains competitive in the low-resource scenario. 
Training i-vector and x-vector extractors on out-of-domain adult speech data lead to comparable performance, where i-vector slightly outperforms x-vector. 
In-domain child speech is preferable in the training of the i-vector system. Using eGeMAPS is less competitive in the detection. 
Given SSD is mainly manifested in phonological errors, the paralinguistic information captured by eGeMAPS may be less informative about the contrast between TD and disordered speech. 

Using speech attribute LPR in the i-vector extraction achieves the best detection performance, followed by phone LPR.
The speech attribute LPR describes multiple changes in articulation and aspiration status. It contains richer diagnostic information than the phone LPR.
Reducing the dimensionality of i-vector with linear discriminant analysis (LDA) further improves the detection and reduces the mis-classification of disordered speakers, giving a UAR of $78.2$\% and a macro F1 score of $78.0$\%.

The results of the best performed i-vector system are analysed using a two-tailed z-test.  A significant threshold of 0.05 is assumed.
The mis-classified subjects are found to be significantly younger than correctly classified subjects. Given word-level ground-truth annotations, the average number of errors in the false negative subjects is significantly less than in the true positive subjects. 
From the clinical perspective, TD children of younger age, i.e. about 3-4, are allowed to make more mistakes than older children in the articulation test. The development of their phonemic inventories and oral motor skills has just begun.
The ambiguity caused by erroneous pronunciations in young TD and disordered subjects makes SSD detection more challenging.

\begin{table}[t!]
\centering
\caption{Performance of subject-level SSD detection based on aggregating word-/phone-level detection results.}
\resizebox{\linewidth}{!}{
\begin{tabular}{|c|c|c|c|c|} 
\hline
\multirow{2}{*}{\textbf{Method}}                                    & \multicolumn{2}{c|}{\textbf{Classifier}}        & \multirow{2}{*}{\textbf{UAR}} & \multirow{2}{*}{\textbf{Macro F1}}  \\ 
\cline{2-3}
                                                                    & Segment             & Subject                   &                               &                                     \\ 
\hline
eGeMAPs                                                             & \multirow{3}{*}{LR} & \multirow{3}{*}{Majority} & 59.4 (8.0)                    & 58.5 (9.5)                          \\ 
\cline{1-1}\cline{4-5}
x-vector                                                             &                     &                           & 63.7 (5.6)                    & 63.7 (6.0)                          \\ 
\cline{1-1}\cline{4-5}
i-vector                                                            &                     &                           & 65.6 (1.9)                    & 65.9 (2.3)                          \\ 
\hline\hline
\begin{tabular}[c]{@{}c@{}}Consonant-vowel \\embedding\end{tabular} & LR                  & SVM                       & \textbf{\textbf{75.6 (4.3)}}  & \textbf{\textbf{72.6 (3.9)}}        \\
\hline
\end{tabular}
}
\label{word-level}
\end{table}

As a comparison to the proposed detection approach, results of subject-level SSD detection based on aggregating phone-level and word-level detection results are given in Table \ref{word-level}. 
Aggregating phone-level detection results from consonant-vowel embeddings deliver competitive performance. 
The detection performance with a UAR of $75.6$\% and a macro F1 score of $72.6$\% is the best among detection methods based on word-level and phone-level results. The performance is also on par with the i-vector approach using MFCCs and utterances of multiple words. Information about context-dependency in consonant errors is effective for detection. 

For SSD detection based on word-level results, 
the i-vector approach delivers a satisfactory performance with $65.6$\% in UAR and $65.9$\% in macro F1 score. 
$62\pm18$\% of the word utterances are classified as correct pronunciations in the disordered speakers. 
$18.4\pm14.6$\% of the word utterances are classified as erroneous in TD speakers. 
The results echo that 
disordered children can correctly produce a significant portion of the test words. 
TD children are also allowed to make age-appropriate errors. 
In the training of classifiers for word-level SSD detection, it would be sub-optimal to assume all word utterances from a speaker are errorless or related to SSD symptoms. 


\section{Conclusion}\label{conclusions}

In this study, we demonstrate the use of knowledge-driven posterior-based features in subject-level SSD detection based on speaker representations. It has been shown that using fixed-dimensional speaker representation extracted from an aggregation of test words is a feasible and effective approach to detecting SSD in child speech. The proposed approach surpasses traditional methods that  combine fine-grained classification results on individual target phones/words. In the low-resource scenario, the i-vector approach outperforms the x-vector one and paralinguistic features. For disorder symptoms that are mainly manifested in phonological errors, extraction of i-vector using posterior-based features can improve detection performance over conventional acoustic features. Our future work will focus on using self-supervised representation learning and neural confidence measure for subject-level SSD detection.


\bibliographystyle{IEEEtran}

\bibliography{mybib}

\end{document}